\documentclass{article}

\usepackage[english]{babel}
\usepackage[a4paper,top=2cm,bottom=2cm,left=3cm,right=3cm,marginparwidth=1.75cm]{geometry}
\usepackage{amssymb}
\usepackage{amsmath}
\usepackage{braket}
\usepackage{siunitx}
\PassOptionsToPackage{hyphens}{url}\usepackage{hyperref}
\usepackage{cleveref}
\usepackage[utf8]{inputenc}
\usepackage[right]{lineno}
\usepackage{csquotes}
\usepackage{booktabs}
\usepackage{longtable}
\usepackage{adjustbox}
\usepackage{array}
\usepackage{url}
\usepackage{titlesec}
\usepackage{authblk}
\usepackage{xcolor}

\titleformat{\subsection}
  {\mdseries\itshape\large}
  {\thesubsection}{1em}{}

\usepackage{natbib}
\crefformat{figure}{#2Figure~#1#3}
\Crefformat{figure}{#2Figure~#1#3}
\crefformat{table}{#2Table~#1#3}
\Crefformat{table}{#2Table~#1#3}
\crefformat{section}{#2Section~#1#3}
\Crefformat{section}{#2Section~#1#3}

\author{Mohammed Hassan}
\author{Omar Abouelazm}

\affil{University of Texas, Department of Computer Science, Austin, Texas}

\title{Improving BB84 Efficiency with Delayed Measurement via Quantum Memory}
\date{}

\begin{document}
\maketitle

\begin{abstract}
In this paper, we introduce a novel modification to the BB84 Quantum Key Distribution (QKD) protocol, aimed at enhancing its efficiency through the use of quantum memory and delayed measurement. In the standard BB84 protocol, the receiver immediately measures the qubits sent by the sender using randomly chosen bases. Due to mismatches between the sender and receiver's bases, a significant portion of the qubits are discarded, reducing the overall key generation rate. Our proposed protocol allows the receiver to store the received qubits in quantum memory and defer measurement until after the sender reveals her basis choices, effectively eliminating the need to discard mismatched qubits.

This modification improves the key generation efficiency while maintaining the core security features of the standard BB84 protocol. By avoiding the unnecessary loss of qubits, our protocol achieves a higher secret key rate without introducing additional vulnerabilities. We present a detailed step-by-step explanation of the delayed measurement process. Although this approach does not alter the security guarantees of BB84, it represents a significant improvement in efficiency, making the protocol more viable for large-scale quantum communication networks.
\end{abstract}

\section{Introduction}

Quantum Key Distribution (QKD) provides a secure method for two distant parties to establish a shared secret key, based on the principles of quantum mechanics. Among the various QKD protocols, the \textit{BB84 protocol}, introduced by Charles Bennett and Gilles Brassard in 1984, is the most widely known and implemented. BB84 ensures security through the quantum mechanical properties of qubits, such as the \textit{no-cloning theorem}, and the fact that any attempt to measure or eavesdrop on quantum states disturbs them, introducing detectable errors \citep{bennett1984quantum, wootters1982single}.

However, despite its robustness, the standard BB84 protocol suffers from inefficiency due to basis mismatches between the sender’s encoding and the receiver’s measurements. In the BB84 protocol, both the sender and receiver randomly select one of two measurement bases. When their choices do not match, the qubits must be discarded, resulting in a significant loss of data—around 50\% of the qubits transmitted cannot be used for key generation \citep{scarani2009security, gisin2002quantum}.

In this paper, we propose a modification to the BB84 protocol that leverages quantum memory and delayed measurement to address this inefficiency. The receiver stores qubits upon receipt rather than measuring them immediately, and measures them only after receiving the sender's basis information. This modification eliminates the need to discard mismatched qubits, increasing the efficiency and overall key generation rate while maintaining the security guarantees of the original protocol.

\section{Related Work}
Research on improving the BB84 protocol has focused on both efficiency and security. Several variations have been proposed, such as the \textit{decoy-state method}, which introduces additional states to detect eavesdropping more accurately, and protocols designed to mitigate source imperfections, like Trojan-horse attacks and mode dependencies \citep{lo2005decoy, zhao2008quantum}.

Quantum memory has also been explored in various contexts related to \textit{QKD}. For instance, in long-distance quantum communication, quantum repeaters rely on quantum memory to extend the range of QKD systems. However, these implementations typically focus on improving communication over large distances rather than addressing the efficiency issues inherent in the standard BB84 protocol \citep{sangouard2011quantum}.

Recent work has introduced several modifications to BB84 that reduce information loss, such as protocols that optimize the selection of measurement bases or use pre-shared information to limit the number of discarded qubits \citep{renner2005information}. However, no work has yet fully explored the combination of delayed measurement and quantum memory to systematically address the problem of basis mismatches in the BB84 protocol.

\section{Proposed Protocol}
In the modified BB84 protocol, the sender transmits qubits as in the original BB84 protocol, using randomly chosen bases. However, instead of immediately measuring the qubits upon receipt, the receiver stores them in a quantum memory. After the transmission of all qubits, the sender communicates the basis used for each qubit through a classical channel. The receiver then measures the stored qubits using the correct basis, thus eliminating the need to discard qubits due to mismatches.

\subsection{BB84 Scenario}
Consider a sender, Alice, receiver, Bob, and an eavesdropper, Eve. Using the standard BB84 protocol, Alice would generate two n-bit binary strings, call them \textit{x} and \textit{y}. She can then generate her qubit string, call it $\ket{\psi}$, by following this table:

\begin{table}[h]
\centering
\caption{Alice's Qubit Generation}
\label{tab:qubit_generation}
\begin{tabular}{ccc}
\hline
\( x_i \) & \( y_i \) & \( |\psi_i\rangle \) \\ \hline
0         & 0         & \( \ket{0} \)       \\
0         & 1         & \( \ket{1} \)       \\
1         & 0         & \( \ket{+} \)       \\
1         & 1         & \( \ket{-} \)       \\ \hline
\end{tabular}
\end{table}

As seen in the chart, the value of the \( x_{i}\) bit determines the correct basis to measure the $\ket{\psi_{i}}$ qubit. If the \( x_{i}\) bit was 0, then Bob needs to measure the $\ket{\psi_{i}}$ qubit in the ($\ket{0}, \ket{1}$) basis, and vice versa.

In order to measure immediately, as necessitated by the standard BB84 protocol, Bob would need to determine what basis to measure each qubit in before Alice sends $\ket{\psi}$. He does so by equivalently generating his own binary string, \textit{x'}. Since \textit{x} and \textit{x'} are binary strings, the number of bits that Bob generated correctly, i.e., \( x'_{i} = x_{i} \), is, on average, \( \frac{n}{2} \).

After measuring $\ket{\psi}$, Bob will now generate his own \textit{y'}. If Eve had not measured the qubits during the transmission, the qubits would not be expected to change, and so the number of \textit{y} bits that Bob generated correctly, i.e., \( y'_{i} = y_{i} \), is, on average, \( \frac{n}{2} \).

Now, Alice and Bob have a shared secret that is, on average, \textbf{length \( \frac{n}{2} \)}.

\subsection{Proposed Protocol Scenario}
Now, consider the proposed modification to the protocol. 
Bob delays his measurement and stores $\ket{\psi}$ in memory, and sends a confirmation-of-receipt bit to Alice. Alice then sends him her basis-determining binary string, \textit{x}. Upon receiving \textit{x}, Bob measures $\ket{\psi}$ in the correct bases.

As before, if Eve had not measured the qubits during the transmission, the qubits would not be expected to change, and so the number of \textit{y} bits that Bob generated correctly, i.e., \( y'_{i} = y_{i} \), is, on average, \textit{n}.

Now, Alice and Bob have a shared secret that is, on average, \textbf{length \textit{n}}.

\subsection{Efficiency Comparison}
This method increases the \textit{efficiency} of the BB84 protocol by a factor of 2, ensuring that a greater percentage of qubits contribute to the final key and enhancing the overall \textit{key generation rate}. The use of \textit{quantum memory} is critical here, as it allows the qubits to be stored without loss of information until the appropriate basis is revealed.

\section{Security Analysis}

\subsection{Review of BB84 Security Principles}

In the standard BB84 protocol, the security is founded on the principles of quantum mechanics, particularly the no-cloning theorem~\citep{wootters1982single}. Alice encodes classical bits into qubits using one of two bases, and Bob measures each qubit in a randomly chosen basis. Since Bob's basis choices are independent of Alice's, there is a 50\% chance that his measurement basis matches Alice's encoding basis for each qubit.

An eavesdropper, Eve, attempting to intercept and measure the qubits will inevitably introduce disturbances due to the no-cloning theorem, which prevents perfect copying of an unknown quantum state~\citep{wootters1982single}. Any measurement by Eve collapses the qubit's state, potentially altering the outcome when Bob measures it. By publicly comparing a subset of their bits, Alice and Bob can estimate the error rate in the transmission~\citep{bennett1984quantum}. A higher-than-expected error rate indicates the presence of an eavesdropper, prompting them to abort the protocol.

\subsection{Security Implications of Storing Qubits in Quantum Memory}

The primary modification in our protocol is Bob's use of quantum memory to store qubits until he receives Alice's basis string \( x \). This change raises several security considerations:

\begin{itemize}
    \item \textbf{Delayed Measurement Vulnerability}: By delaying measurement until after basis reconciliation, there is a concern that an eavesdropper could exploit this window to gain information without detection~\citep{gisin2002quantum}. However, the security against such an attack relies on the integrity and isolation of Bob's quantum memory. If Eve cannot access or tamper with the stored qubits, the security remains intact.
    \item \textbf{Quantum Memory Imperfections}: Practical quantum memories are subject to decoherence and operational errors~\citep{lvovsky2009optical}, which could introduce additional noise. This noise could mask an eavesdropper's presence by increasing the baseline error rate. To mitigate this, the quantum memory's error rates must be well-characterized and sufficiently low to distinguish between intrinsic memory errors and those introduced by eavesdropping.
    \item \textbf{Basis Revelation Timing}: Since Alice sends her basis string \( x \) after Bob confirms receipt of the qubits, there is a potential risk if Eve can intercept both the qubits and the basis information. To prevent this, the classical communication channels must be authenticated and secured against man-in-the-middle attacks, as emphasized in the original BB84 protocol~\citep{bennett1984quantum}.
\end{itemize}

\subsection{Maintaining Security Through Authentication and Error Correction}

To ensure the modified protocol's security aligns with that of the standard BB84, we incorporate the following measures:

\begin{itemize}
    \item \textbf{Authenticated Classical Channels}: As in the original BB84, the classical communication used for basis reconciliation and error rate estimation must be authenticated~\citep{mayers2001unconditional}. This prevents Eve from impersonating Alice or Bob and injecting false information.
    \item \textbf{Error Rate Monitoring}: After Bob measures the qubits using Alice's basis string \( x \), they perform error rate estimation by comparing a subset of their bit strings. This process is identical to the standard BB84 and allows them to detect any eavesdropping attempts that introduce detectable disturbances~\citep{bennett1984quantum}.
    \item \textbf{Privacy Amplification and Error Correction}: Even if Eve gains partial information about the key, privacy amplification techniques can distill a secure key from the partially compromised one~\citep{bennett1988privacy}. This is a direct application of the methods used in the original BB84 protocol.
\end{itemize}

By ensuring that the quantum memory is secure and that all classical communications are authenticated, the modified protocol retains the security guarantees provided by the original BB84 protocol. The key difference lies in operational efficiency rather than security fundamentals.

\section{Practical Considerations}

While the proposed protocol offers significant efficiency gains by reducing the number of discarded qubits, its practical implementation hinges on advancements in quantum memory technology. Current quantum memory systems, although promising, still face substantial challenges related to decoherence, limited storage times, and operational fidelity~\citep{sangouard2011quantum, lvovsky2009optical}. Decoherence leads to the loss of quantum information over time, introducing errors that can compromise the protocol's security and efficiency. However, with rapid developments in this field, it is reasonable to anticipate that quantum memories with longer coherence times and higher fidelity will become viable in the near future~\citep{heshami2016quantum, bussieres2013prospective}.

Furthermore, the modified protocol may introduce practical complexities related to timing and synchronization. Storing qubits in quantum memory until the basis information is received requires precise control over the qubits' coherence times and synchronization between Alice and Bob. Implementing this level of control can be technically challenging and may necessitate sophisticated hardware and protocols~\citep{lvovsky2009optical, heshami2016quantum}.

Despite these challenges, the potential benefits of the modified protocol justify the pursuit of practical solutions. The reduction in qubit wastage and the increased efficiency in key generation could have significant implications for the scalability and practicality of quantum key distribution systems. Continued research and development in quantum memory technologies and secure classical communication protocols are essential steps toward realizing the full potential of this modified BB84 protocol.

\section{Conclusion}

In this paper, we have proposed a novel modification to the BB84 quantum key distribution protocol that leverages quantum memory to delay the measurement of qubits until after the basis information has been shared. This approach effectively eliminates the need to discard qubits due to basis mismatches, thereby significantly increasing the key generation rate while preserving the fundamental security properties of the original BB84 protocol.

Our security analysis demonstrates that the modified protocol maintains robustness against eavesdropping attempts, as it adheres to the core quantum mechanical principles that underlie the security of BB84, including the no-cloning theorem and the measurement disturbance phenomenon. By storing qubits in quantum memory, Bob can measure all received qubits in the correct bases, enhancing efficiency without compromising security.

We have also addressed practical considerations, acknowledging that the implementation of this modified protocol depends on advancements in quantum memory technology. Current limitations such as decoherence and limited storage times present challenges; however, ongoing research in quantum memories shows promising progress toward overcoming these hurdles.

The proposed modification offers a compelling direction for future research in quantum cryptography. By enhancing the efficiency of key generation, it brings us closer to practical, high-rate quantum communication systems. We encourage further exploration into the development of robust quantum memories and the practical integration of such technologies into quantum key distribution protocols.

% \bibliography{Bibliography}

\begin{thebibliography}{13}
\providecommand{\natexlab}[1]{#1}
\providecommand{\url}[1]{\texttt{#1}}
\expandafter\ifx\csname urlstyle\endcsname\relax
  \providecommand{\doi}[1]{doi: #1}\else
  \providecommand{\doi}{doi: \begingroup \urlstyle{rm}\Url}\fi

\bibitem[Bennett and Brassard(1984)]{bennett1984quantum}
Charles~H. Bennett and Gilles Brassard.
\newblock Quantum cryptography: Public key distribution and coin tossing.
\newblock \emph{Proceedings of IEEE International Conference on Computers, Systems and Signal Processing}, pages 175--179, 1984.

\bibitem[Bennett et~al.(1988)Bennett, Brassard, and Robert]{bennett1988privacy}
Charles~H. Bennett, Gilles Brassard, and Jean-Marc Robert.
\newblock Privacy amplification by public discussion.
\newblock \emph{SIAM Journal on Computing}, 17\penalty0 (2):\penalty0 210--229, 1988.

\bibitem[Bussi{\`e}res et~al.(2013)Bussi{\`e}res, Sangouard, Afzelius, de~Riedmatten, Simon, and Tittel]{bussieres2013prospective}
F{\'e}lix Bussi{\`e}res, Nicolas Sangouard, Mikael Afzelius, Hugues de~Riedmatten, Christoph Simon, and Wolfgang Tittel.
\newblock Prospective applications of optical quantum memories.
\newblock \emph{Journal of Modern Optics}, 60\penalty0 (18):\penalty0 1519--1537, 2013.

\bibitem[Gisin et~al.(2002)Gisin, Ribordy, Tittel, and Zbinden]{gisin2002quantum}
Nicolas Gisin, Gr{\'e}goire Ribordy, Wolfgang Tittel, and Hugo Zbinden.
\newblock Quantum cryptography.
\newblock \emph{Reviews of Modern Physics}, 74\penalty0 (1):\penalty0 145--195, 2002.

\bibitem[Heshami et~al.(2016)Heshami, England, Humphreys, Bustard, Acosta, Nunn, and Sussman]{heshami2016quantum}
Khabat Heshami, Duncan~G. England, Peter~C. Humphreys, Philip~J. Bustard, Victor~M. Acosta, Joshua Nunn, and Benjamin~J. Sussman.
\newblock Quantum memories: emerging applications and recent advances.
\newblock \emph{Journal of Modern Optics}, 63\penalty0 (20):\penalty0 2005--2028, 2016.

\bibitem[Lo et~al.(2005)Lo, Ma, and Chen]{lo2005decoy}
Hoi-Kwong Lo, Xiongfeng Ma, and Kai Chen.
\newblock Decoy state quantum key distribution.
\newblock \emph{Physical Review Letters}, 94\penalty0 (23):\penalty0 230504, 2005.

\bibitem[Lvovsky et~al.(2009)Lvovsky, Sanders, and Tittel]{lvovsky2009optical}
Alexander~I. Lvovsky, Barry~C. Sanders, and Wolfgang Tittel.
\newblock Optical quantum memory.
\newblock \emph{Nature Photonics}, 3\penalty0 (12):\penalty0 706--714, 2009.

\bibitem[Mayers(2001)]{mayers2001unconditional}
Dominic Mayers.
\newblock Unconditional security in quantum cryptography.
\newblock \emph{Journal of the ACM}, 48\penalty0 (3):\penalty0 351--406, 2001.

\bibitem[Renner and K{\"o}nig(2005)]{renner2005information}
Renato Renner and Robert K{\"o}nig.
\newblock Information-theoretic security proof for quantum-key-distribution protocols.
\newblock \emph{Physical Review Letters}, 95\penalty0 (8):\penalty0 080501, 2005.

\bibitem[Sangouard et~al.(2011)Sangouard, Simon, de~Riedmatten, and Gisin]{sangouard2011quantum}
Nicolas Sangouard, Christoph Simon, Hugues de~Riedmatten, and Nicolas Gisin.
\newblock Quantum repeaters based on atomic ensembles and linear optics.
\newblock \emph{Reviews of Modern Physics}, 83\penalty0 (1):\penalty0 33--80, 2011.

\bibitem[Scarani et~al.(2009)Scarani, Bechmann-Pasquinucci, Cerf, Du{\v{s}}ek, L{\"u}tkenhaus, and Peev]{scarani2009security}
Valerio Scarani, Helle Bechmann-Pasquinucci, Nicolas~J. Cerf, Miloslav Du{\v{s}}ek, Norbert L{\"u}tkenhaus, and Momtchil Peev.
\newblock The security of practical quantum key distribution.
\newblock \emph{Reviews of Modern Physics}, 81\penalty0 (3):\penalty0 1301--1350, 2009.

\bibitem[Wootters and Zurek(1982)]{wootters1982single}
William~K. Wootters and Wojciech~H. Zurek.
\newblock A single quantum cannot be cloned.
\newblock \emph{Nature}, 299\penalty0 (5886):\penalty0 802--803, 1982.

\bibitem[Zhao et~al.(2008)Zhao, Qi, Ma, Lo, and Qian]{zhao2008quantum}
Yi~Zhao, Bing Qi, Xiongfeng Ma, Hoi-Kwong Lo, and Li~Qian.
\newblock Quantum key distribution with imperfect devices.
\newblock \emph{Physical Review A}, 78\penalty0 (4):\penalty0 042333, 2008.

\end{thebibliography}

% Include the .bbl file directly

\end{document}